\documentclass[aps,preprint,nofootinbib]{revtex4}%
\usepackage{amsfonts}
\usepackage{amsmath}
\usepackage{amssymb}
\usepackage{graphicx}%
\providecommand{\U}[1]{\protect\rule{.1in}{.1in}}

\begin{document}
\preprint{ }
\title[Short title for running header]{Lagrangian symmetries of the ADM action. Do we need a solution to the
\textquotedblleft non-canonicity puzzle\textquotedblright?}
\author{N. Kiriushcheva}
\email{nkiriush@uwo.ca}
\author{P. G. Komorowski}
\email{pkomoro@uwo.ca}
\author{S. V. Kuzmin}
\email{skuzmin@uwo.ca}
\affiliation{The Department of Applied Mathematics, The University of Western Ontario,
London, Ontario, Canada, N6A 5B7}
\pacs{PACS number}

\begin{abstract}
We argue that there is nothing puzzling in the fact that the Hamiltonian
formulation of a covariant theory, General Relativity, after a non-covariant
change of field variables is not canonically related to the formulation based
on the original variable, the metric tensor. Were such a puzzle to be
\textquotedblleft solved\textquotedblright\ it would lead to the conclusion
that a covariant theory can be converted into a non-covariant one in many
different ways and without consequence. The non-canonicity of transformations
from covariant to non-covariant variables shows the need to work in the
original variables so as to be able to restore the covariant gauge
transformations in the Hamiltonian approach. Any modification of Dirac's
procedure for the constrained Hamiltonian with the aim to prove the legitimacy
of non-covariant changes of field variables, or rejection of Dirac's procedure
as \textquotedblleft not fundamental and undoubted\textquotedblright\ because
it does not allow such changes (as recently suggested by Shestakova
\textit{[CGQ, 28 (2011) 055009]}), is equivalent to the rejection of the
covariance of General Relativity, and to surrender to such temptation is truly puzzling.

\end{abstract}
\volumeyear{year}
\volumenumber{number}
\issuenumber{number}
\eid{identifier}
\date{\today}
\received{}

\maketitle


\hspace{5cm}\begin{minipage}{10cm}
\begin{flushright}
\textit{\textquotedblleft Numquam ponenda est pluralitas sine necessitate.\textquotedblright}
\\
William of Ockham (1285-1349) \\
\textit{ Plurality must never be posited without necessity.}
\end{flushright}
\end{minipage}

\vspace{1cm}

\section{Introduction}

The term \textquotedblleft non-canonicity puzzle\textquotedblright\ was coined
in \cite{CLM} to describe the apparently contradictory results of applying the
Dirac procedure to the Hamiltonian formulations of constrained systems
\cite{Diracbook}, in particular, to the Einstein-Hilbert (EH) and to the
Arnowitt-Deser-Misner (ADM) \cite{ADM, goldies} actions. The contradiction
arises from the fact that the ADM action can be obtained from the EH action by
an invertible change of variables, which is an admissible procedure at the
Lagrangian level. Yet at the same time, the corresponding Hamiltonians are not
related to each other by a canonical transformation \cite{Myths}; therefore,
they are not equivalent at the Hamiltonian level. We shall use Lagrangian
methods to explore this apparent paradox.

In the case of the two older Hamiltonian formulations due to Pirani, Schild,
and Skinner (PSS) \cite{PSS} and due to Dirac \cite{Dirac}, where the metric
tensor is used as a field variable, the first-class constraints and their
algebra of Poisson brackets (PBs) lead to diffeomorphism
invariance\footnote{We understand diffeomorphism invariance (\textit{diff}) to
mean \textquotedblleft active\textquotedblright\ \cite{Rovelli} (p. 62), when
\textquotedblleft coordinates play no role\textquotedblright, i.e. the
transformations of fields are written in the same coordinate system.} in both
formulations, as demonstrated in \cite{KKRV} and \cite{Myths}. The PSS and
Dirac Lagragians differ by a total derivative that Dirac added to simplify the
primary constraints \cite{Dirac}; this difference does not affect the
equations of motion, which in both cases are Einstein's original equations, as
the field variables in these two formulations are the same: the components of
the metric tensor. When the same Dirac Hamiltonian method is applied to the
ADM action, a different (though unique) symmetry follows, which is known by
many names: \textquotedblleft spatial diffeomorphism\textquotedblright%
,\ \textquotedblleft special induced diffeomorphism\textquotedblright%
,\ \textquotedblleft field-dependent diffeomorphism\textquotedblright%
,\ \textquotedblleft foliation preserving diffeomorphism\textquotedblright,
\textquotedblleft one-to-one correspondence\textquotedblright, and
\textquotedblleft one-to-one mapping\textquotedblright\ (see \cite{Myths} and
references therein). Because the two symmetries, diffeomorphism and the
symmetry that follows from the Hamiltonian analysis of the ADM action, are
distinct, it is no surprise that the corresponding Hamiltonians are not
equivalent and that the field variables are not related by a canonical
transformation, as was explicitly demonstrated in \cite{Myths}. Further, as it
was conjectured in \cite{FKK} for Hamiltonians with constraints, based on the
comparison of the two equivalent Hamiltonians (leading to the same symmetry)
of PSS \cite{PSS, KKRV} and Dirac \cite{Dirac, Myths}, the ordinary canonicity
condition \cite{Lanczosbook} (PBs for two sets of phase-space variables) is
only a necessary condition; the whole structure of PB algebra of first-class
constraints must also be preserved.

There is nothing puzzling in the non-equivalence of two Hamiltonians with
different symmetries. But the change of field variables performed by ADM is
invertible, and such changes could keep two actions equivalent; yet, in the
case of the EH action we are dealing with a singular (gauge invariant) and
generally covariant theory, and in passing to the ADM action, a non-linear and
non-covariant change of variables is performed. How does this change affect
the results? The disappearance of diffeomorphism invariance in the Hamiltonian
formulation of the ADM action was proclaimed long ago, e.g. in the statement
of Isham and Kuchar \cite{Isham}: \textquotedblleft the full group of
spacetime diffeomorphism has \textit{somehow} got lost in making the
transition from the Hilbert action to the Dirac-ADM action\textquotedblright%
\ (italic is ours)\footnote{The name of Dirac is used incorrectly in this
statement because the Dirac Hamiltonian is not canonically related to the ADM
Hamiltonian \cite{Myths}. In addition, Dirac's modification of the EH action
is performed in a way that preserves Einstein's equations \cite{Dirac}.}. If
diffeomorphism is the gauge symmetry of the EH action and its Hamiltonian
formulation, and it \textquotedblleft got lost\textquotedblright\ in the ADM
action, then the two actions cannot be equivalent; and the \textquotedblleft%
\textit{somehow}\textquotedblright\ only arises in making the transition
\textit{--} a non-covariant and non-linear change of field variables. To us,
it appears that the conclusion in \cite{Isham} was based on the results of
Hamiltonian analysis, not on an analysis at the Lagrangian level, about which
\textquotedblleft the ADM action\textquotedblright\ statement is made.

In \cite{KKK-1} we considered the symmetries of the EH action at the
Lagrangian level. We compared the diffeomorphism transformation and the
transformation that follows from the Hamiltonian analysis of the ADM action
for the same field: the metric tensor (the ADM transformations are known, and
the redefinition of their variables in terms of the metric can be used to
calculate how the metric is transformed under this symmetry). The ADM
transformations of the metric tensor can be formally presented as
diffeomorphism invariance with field-dependent gauge parameters (this is the
origin of some names for the ADM transformations, e.g. \textquotedblleft
field-dependent diffeomorphism\textquotedblright). At the Lagrangian level, by
Noether's second theorem \cite{Noether, Noether-eng}, gauge symmetries are
related to differential identities (DIs) \textit{--} combinations of
Euler-Lagrange derivatives (ELDs) that are identically zero (off-shell). Even
if some symmetry could be formally presented as another symmetry by a
field-dependent redefinition of gauge parameters, these symmetries would be
distinct since they correspond to different DIs (see \cite{KKK-1}). A new DI,
obtained in such a way, is an identity; thus, by the converse of Noether's
second theorem, the ADM transformations of the metric tensor, which are
described by such DIs, are also a symmetry of the EH action; and many other
\textquotedblleft field-dependent diffeomorphisms\textquotedblright\ can be
constructed by using different redefinitions of the parameters. Is such a
plurality of symmetries also a plurality of equally good outcomes? The study
of the group properties of such transformations for the EH action \cite{KKK-1}
shows that only one (unique) symmetry has group properties: diffeomorphism;
and the other symmetries, which are easily constructed by relating the gauge
symmetries by using a field-dependent redefinition of the gauge parameters,
including the ADM symmetry, do not have group properties. When the Dirac
algorithm is applied to the EH action it leads exactly to this one (unique),
or shall we say, canonical, gauge symmetry of the EH action (there is no
puzzle), and the metric is a canonical variable of the canonical Hamiltonian
formulation of General Relativity (PSS and Dirac). Note that the DI that
describes the diffeomorphism of the metric tensor is a covariant derivative of
a true tensor density; consequently, it is identically zero in all coordinate
systems. But all DIs obtained by a non-covariant field-dependent redefinition
of gauge parameters are not true tensors; therefore, the identities are not
valid in all coordinate systems.

In this paper we continue the investigation we began in \cite{KKK-1} by
studying different symmetries and their group properties for the ADM action.
In the next Section we briefly review the origin of the ADM Lagrangian and
establish notation. In Section III we study the group properties of the
symmetry, which follows from applying Dirac's Hamiltonian method to the ADM
action. In Section IV we consider the invariance of the ADM action under
diffeomorphism (written for the ADM variables) and its group property. In
Discussion we analyze the results of the Lagrangian consideration for the EH
and ADM actions, and discuss their relation to the Dirac Hamiltonian analysis
of the same actions. We summarize this work in Conclusion.

\section{Prelude. The ADM Lagrangian}

A long chain of manipulations performed on the EH Lagrangian density gives
rise to the ADM Lagrangian\footnote{We use Greek letters for space-time
indices, $\mu=0,1,2,3$, and Latin letters for space indices, $i=1,2,3$.},%

\begin{equation}
L_{EH}\left(  g_{\mu\nu}\right)  \longrightarrow L_{\Gamma\Gamma}\left(
g_{\mu\nu}\right)  \longrightarrow L_{Dirac}\left(  g_{\mu\nu}\right)
\longrightarrow L_{ADM}\left(  N,N^{i},\gamma_{km}\right)  . \label{eqn1}%
\end{equation}
It is only for the ADM Lagrangian that the original field variable, the metric
tensor, has been changed to new, non-covariant variables.

The EH action, written in manifestly covariant form, is \cite{Landau, Carmeli}%

\begin{equation}
S_{EH}\left(  g_{\mu\nu}\right)  =\int L_{EH}\left(  g_{\mu\nu}\right)
d^{4}x=\int\sqrt{-g}Rd^{4}x~, \label{eqn2}%
\end{equation}
where the Ricci scalar, $R$, contains terms with second-order derivatives of
the metric. The presence of second-order derivatives in time
(\textquotedblleft accelerations\textquotedblright) does not allow one to pass
to the Hamiltonian formulation directly, since momenta have to be introduced
by performing a variation of the Lagrangian with respect to the first-order
derivatives in time of the metric (the \textquotedblleft
velocities\textquotedblright). Despite the belief that \textquotedblleft
accelerations\textquotedblright\ in the action forbid \textquotedblleft any
canonical treatment of the theory\textquotedblright\ \cite{CLM}, it is
possible to work with the original action (where \textquotedblleft
accelerations\textquotedblright\ are present) by a generalization of the
Ostrogradsky method \cite{Ostr, Ostr-rus} (see also \cite{Whittaker}) and to
find Hamiltonians for Lagrangians with higher order derivatives, after a
proper generalization of the Ostrogradsky method for constrained systems
\cite{GLT} (also see \cite{GTbook}). An attempt to work with (\ref{eqn2}) was
made in \cite{DD}, but to the best of our knowledge it was never completed,
although in our opinion it should be possible. An example of how to perform
the Hamiltonian analysis using the Ostrogradsky method was given in \cite{2D}
for the two-dimensional metric and tetrad gravities.

In the first Hamiltonian formulation of the EH action \cite{PSS} the need to
work with the second-order derivatives in (\ref{eqn2}) was avoided by using
the so-called gamma-gamma form (without second-order derivatives of the metric
in the Lagrangian), which leads to the same equations of motion,%

\begin{equation}
S_{\Gamma\Gamma}\left(  g_{\mu\nu}\right)  =\int L_{\Gamma\Gamma}\left(
g_{\mu\nu}\right)  d^{4}x=\int\sqrt{-g}g^{\mu\nu}\left(  \Gamma_{\mu\nu}%
^{\rho}\Gamma_{\rho\sigma}^{\sigma}-\Gamma_{\mu\rho}^{\sigma}\Gamma_{\nu
\sigma}^{\rho}\right)  d^{4}x~. \label{eqn3}%
\end{equation}

Action (\ref{eqn3}) is quadratic in the first-order derivatives of the metric,
so it is at most quadratic in \textquotedblleft velocities\textquotedblright;
because of this, it is well suited for standard Hamiltonian formulations. The
Hamiltonian formulation of the EH action in the gamma-gamma form, using the
Dirac procedure, was outlined in \cite{PSS} but not completed; at that time
the technique to restore gauge invariance from the first-class constraints was
not known, and even Dirac's conjecture about the connection of the first-class
constraints and gauge transformations appeared much later \cite{Diracbook}.
From the algebra of constraints of the PSS formulation \cite{PSS}, the
restoration of gauge symmetries was performed in \cite{KKRV} by using the
Castellani procedure \cite{Castellani}. The total Hamiltonian has the form
\cite{KKRV}%

\begin{equation}
H_{T}=\dot{g}_{0\mu}\phi^{0\mu}+g_{0\mu}\chi_{PSS}^{0\mu}~, \label{eqn4}%
\end{equation}
and the entire set of first-class constraints ($\phi^{0\mu}$\textit{--}
primary and $\chi^{0\mu}$\textit{--} secondary) leads to diffeomorphism
invariance \cite{KKRV}.

Without affecting the equations of motion or changing the original variables,
the next modification was performed a few years later by Dirac \cite{Dirac}
who added two total derivatives to the Lagrangian $L_{\Gamma\Gamma}$: \ %

\begin{equation}
S_{Dirac}\left(  g_{\mu\nu}\right)  =\int L_{Dirac}\left(  g_{\mu\nu}\right)
d^{4}x \label{eqn4a}%
\end{equation}

\[
=\int\left\{  L_{\Gamma\Gamma}\left(  g_{\mu\nu}\right)  +\left[  \left(
\sqrt{-g}g^{00}\right)  _{,v}\frac{g^{v0}}{g^{00}}\right]  _{,0}-\left[
\left(  \sqrt{-g}g^{00}\right)  _{,0}\frac{g^{v0}}{g^{00}}\right]
_{,v}\right\}  d^{4}x~.
\]

The main goal of Dirac's modification was to simplify the primary constraints
of \cite{PSS}, which are not pure momenta for the $S_{\Gamma\Gamma}$ action.
In addition, some rearrangements were made in the course of calculating the
Hamiltonian \cite{Dirac}. (The detail analysis of Dirac's paper can be found
in \cite{Myths}.) The total Hamiltonian of Dirac is%

\begin{equation}
H_{T}=\dot{g}_{0\mu}\pi^{0\mu}+g_{0\mu}\chi_{Dirac}^{0\mu}~. \label{eqn5}%
\end{equation}

Note that the primary and secondary first-class constraints in (\ref{eqn4})
and (\ref{eqn5}) have a different form, but the algebra of PBs among the
constraints is the same in both the Dirac and PSS formulations \cite{FKK}. The
gauge transformations that correspond to the Dirac constraints were calculated
in \cite{Myths}, and as in the case of the PSS formulation, they also lead to
diffeomorphism invariance. Comparison of the two Hamiltonians (\ref{eqn4}) and
(\ref{eqn5}) revealed that their phase-space variables are related by a
canonical transformation \cite{FKK} (not surprisingly, as they have the same
symmetry), which in addition preserves the structure of the constraint algebra
(although the constraints themselves are different). Both Lagrangians
(\ref{eqn2}) and (\ref{eqn3}) are functionals of the metric tensor, and the
phase-space variables in both Hamiltonians are the metric and corresponding momenta.

The ADM Lagrangian can be obtained by performing a change of field variables
in the final form of the Dirac Lagrangian (which is based on the Hamiltonian
analysis, see Section 4.5 of \cite{Myths}):%

\begin{equation}
N=\left(  -g^{00}\right)  ^{-1/2},\text{~}N^{i}=-\frac{g^{0i}}{g^{00}}%
,~\gamma_{km}=g_{km}~. \label{eqn10}%
\end{equation}
This change of field variables is non-linear, non-covariant, yet invertible.
The names assigned to these variables, lapse $N$ and shift $N^{i}$ functions,
reflect the non-covariant nature of this redefinition. In some works $N_{i}$
is used instead of $N^{i}$ as an independent variable, but of course this
cannot affect the results. Our choice of $N^{i}$ is dictated by a particular
and common expression for the secondary first-class constraints of the ADM
Hamiltonian $N^{\mu}\mathcal{H}_{\mu}$ (e.g. see \cite{Pons2003})%

\begin{equation}
H_{T}=\dot{N}^{\mu}p_{\mu}+N^{\mu}\mathcal{H}_{\mu}~. \label{eqn11}%
\end{equation}

Note that this form is not covariant, as $N^{\mu}=\left(  N,N^{i}\right)  $
and \ $\mathcal{H}_{\mu}=\left(  \mathcal{H}_{\bot},\mathcal{H}_{k}\right)  $
are quantities that are neither vectors nor components of some covariant
quantities; unlike the Hamiltonian formulation of the EH action in the PSS or
Dirac forms, where the secondary constraints $\chi^{0\mu}$ enter the total
Hamiltonian as a combination $g_{0\mu}\chi^{0\mu}$, which is a contraction of
$\chi^{0\mu}$ with the components of a true tensor (see (\ref{eqn5}) and
(\ref{eqn10})). The inverse transformations for the covariant and
contravariant metric tensors are easy to find from (\ref{eqn10}) by using
$g_{\mu\nu}g^{\nu\alpha}=\delta_{\mu}^{\alpha}$ (see also, e.g.
\cite{Castellani}):
\begin{equation}
g_{\mu\nu}=\left(
\begin{array}
[c]{cc}%
\gamma_{ij}N^{i}N^{j}-N^{2} & \gamma_{ij}N^{j}\\
\gamma_{ij}N^{j} & \gamma_{ij}%
\end{array}
\right)  ,~g^{\mu\nu}=\left(
\begin{array}
[c]{cc}%
-\frac{1}{N^{2}} & \frac{N^{i}}{N^{2}}\\
\frac{N^{i}}{N^{2}} & \gamma^{ij}-\frac{N^{i}N^{j}}{N^{2}}%
\end{array}
\right)  ;\label{eqn12}%
\end{equation}
where $\gamma^{ik}$ is defined by $\gamma^{ik}\gamma_{km\text{ }}=\delta
_{m}^{i}$. We note that the independent field variables of the ADM Lagrangian
are $N,N^{i},$ and $g_{km}$, so the combination $\gamma^{km}-\frac{N^{k}N^{m}%
}{N^{2}}$ is $g^{km}$, since $\gamma^{km}=e^{km}=g^{km}-\frac{g^{0k}g^{om}%
}{g^{00}}=$ $g^{km}+\frac{N^{k}N^{m}}{N^{2}}$. The combination $e^{km}$ was
introduced by Dirac \cite{Dirac}; and to avoid unnecessary complications with
notation, we shall use $g_{km}$ and Dirac's short form $e^{km}$, which makes
the comparison of transformations for the metric and the ADM variables more
transparent. The ADM Lagrangian is reported in many places (e.g.
\cite{Gravitation, Waldbook}), and it is usually written in the following
form:
\begin{equation}
L_{ADM}=\sqrt{\det g_{km}}N\left(  E^{rsab}K_{rs}K_{ab}+R_{3}\right)
,\label{eqn15}%
\end{equation}
where%

\begin{equation}
E^{rsab}=e^{rs}e^{ab}-e^{rb}e^{as}, \label{eqn16}%
\end{equation}

\begin{equation}
K_{rs}=\frac{1}{2N}\left(  g_{rp}N_{,s}^{p}+g_{sp}N_{,r}^{p}+N^{p}%
g_{rs,p}-g_{rs,0}\right)  , \label{eqn17}%
\end{equation}

\begin{equation}
R_{3}=g_{mn,kt}E^{mnkt}+g_{mn,k}g_{pq,t}F^{mnkpqt} \label{eqn20}%
\end{equation}
with%

\begin{equation}
F^{mnkpqt}=\frac{1}{4}\left(  E^{mnpq}e^{kt}-2E^{ktpn}e^{mq}-4E^{pqnt}%
e^{mk}\right)  . \label{eqn22}%
\end{equation}

Note that the form of $R_{3}$ is the result of a calculation in the Dirac
Hamiltonian formulation.\ And $R_{3}$ is also equivalent to a similar
expression used by Dutt and Dresden \cite{DD}. The Dirac Lagrangian is not
only a modification of $L_{\Gamma\Gamma}$ of the PSS Lagrangian, but because
of further rearrangements it also contains expressions that include terms with
second-order derivatives, as in (\ref{eqn2}) (for the details of how the ADM
Lagrangian follows from the Dirac Hamiltonian analysis by the change of
variables (\ref{eqn10}) see \cite{Myths}). Equation (\ref{eqn17}) is written
in a form where the basic variables $N,N^{i},$ and $g_{km}$ are presented
explicitly. The shorter, more frequently used form of (\ref{eqn17}) is less
suitable for calculations; but nothing can be shorter than $\sqrt{-g}R$, and
for a covariant theory the covariant form is always preferable.

The Dirac procedure applied to the ADM Lagrangian (\ref{eqn15}) leads to the
total Hamiltonian in the form (\ref{eqn11}). The gauge transformations that
follow are not diffeomorphism, and this formulation is not canonically related
to Dirac's (this is not unexpected since they have different gauge
transformations). This difference of transformations in the Hamiltonian
approach leads to the conclusion (e.g. of Isham and Kuchar) that
\textquotedblleft diffeomorphism has somehow got lost in making the
transition\textquotedblright\ from the EH action to the ADM action\footnote{In
the literature, despite this loss of diffeomorphism, which is the gauge
symmetry of the EH action and Hamiltonian formulations of the PSS and Dirac
actions, the equivalence of EH and ADM actions is presumed, e.g.\ in the title
of the original ADM paper \cite{ADM}: \textquotedblleft Dynamics of General
Relativity\textquotedblright, or in \cite{Pons2003}: \textquotedblleft the
canonical formalism [ADM] of GR\textquotedblright,\ or in \cite{Horava}:
\textquotedblleft the ADM decomposition of the Einstein-Hilbert
action\textquotedblright,\ or in \cite{Pullin}: \textquotedblleft the true
Hamiltonian dynamics of general relativity\textquotedblright.}; and if a gauge
symmetry \textquotedblleft got lost\textquotedblright,\ then two actions
cannot be equivalent. In paper \cite{KKK-1} we considered the invariance of
the EH action under two transformations (diffeomorphism and ADM) and
demonstrated that both, and many others, are symmetries of the EH action; but
the one that has a group property is diffeomorphism. In this paper we perform
the same analysis for the ADM Lagrangian, and for the same two symmetries. We
shall investigate whether diffeomorphism has really \textit{been lost}\ or if
it is a symmetry of the ADM action; and we shall investigate whether or not it
is the only symmetry with a group property, as in the case of the EH action,
or do the ADM transformations for the ADM action\ also have a group property?

\section{Group properties of the ADM transformations}

Use of the standard Dirac procedure\footnote{Dirac's procedure includes:
introducing momenta to all fields; finding the primary constraints;
considering their time development until the closure is reached; eliminating
second-class constraints; constructing gauge generators using all first-class
constraints; and finding the transformations of all phase-space variables,
which after elimination of momenta give the gauge transformations in terms of
the original variables that enter the Lagrangian. For the EH action in the
gamma-gamma form (second order), there are no second-class constraints; but,
e.g. for the affine-metric, first-order, formulation of GR \cite{Einstein1925,
Einstein1925-rus, Einstein-eng}, second-class constraints do appear
\cite{KK-Annals, affine-metric, G/R, Gerry}.} for the Hamiltonian formulation
of the ADM Lagrangian (\ref{eqn11}) leads to the following gauge
transformations (e.g. see appendix of \cite{Castellani} and Section 4.3 of
\cite{Myths}):
\begin{equation}
\delta_{ADM}N=\varepsilon_{,j}^{\bot}N^{j}-\varepsilon_{,0}^{\bot}%
-\varepsilon^{i}N_{,i}~,\label{eqn40}%
\end{equation}

\begin{equation}
\delta_{ADM}N^{k}=-\varepsilon^{\bot}N_{,j}e^{jk}+\varepsilon_{,j}^{\bot
}Ne^{kj}-\varepsilon^{j}N_{,j}^{k}+\varepsilon_{,j}^{k}N^{j}-\varepsilon
_{,0}^{k}~, \label{eqn41}%
\end{equation}
and%

\begin{equation}
\delta_{ADM}g_{km}=\varepsilon^{\bot}\frac{1}{N}\left[  g_{kn}N_{,m}%
^{n}+g_{mn}N_{,k}^{n}-g_{km,0}+N^{n}g_{km,n}\right]  -\varepsilon_{,m}%
^{i}g_{ik}-\varepsilon_{,k}^{i}g_{im}-\varepsilon^{i}g_{km,i}~, \label{eqn42}%
\end{equation}
where $\varepsilon^{\bot}$ and $\varepsilon^{i}$ are gauge (field-independent)
parameters. It would be a heroic task to show directly the invariance of the
ADM action (\ref{eqn15}) under transformations (\ref{eqn40})-(\ref{eqn42}).
Alternatively, we can use Noether's second theorem, which allows one to find a
corresponding identity if a gauge transformation is known. It is a
straightforward procedure to check whether or not an identity that has been
found in this way is satisfied. We will also need the DIs that correspond to
(\ref{eqn40})-(\ref{eqn42}) in the next Section.

Using the prescription outlined in \cite{Schwinger}, we write%

\begin{equation}
\delta S_{ADM}=\int\left[  E\delta N+E_{i}\delta N^{i}+E_{ADM}^{km}\delta
g_{km}\right]  d^{4}x=\int\left[  \varepsilon^{\bot}I_{\bot}^{ADM}%
+\varepsilon^{k}I_{k}^{ADM}\right]  d^{4}x, \label{eqn45}%
\end{equation}
where $E$, $E_{i}$, and $E_{ADM}^{km}$ are the Euler-Lagrange derivatives of
the ADM action, i.e. $E=\frac{\delta L_{ADM}}{\delta N}$, \textit{et cetera.}
(We use the subscript `$ADM$' in $E_{ADM}^{km}$ to distinguish it from the
$E^{km}=\frac{\delta L_{EH}}{\delta g_{km}}$ for the EH action.) The
substitution of transformations (\ref{eqn40})-(\ref{eqn42}) into (\ref{eqn45})
and a rearrangement of terms (including integration by parts) to single out
the gauge parameters lead to the following DIs:%

\[
I_{\bot}^{ADM}=-\left(  N^{j}E\right)  _{,j}+E_{,0}-N_{,j}e^{jk}E_{k}-\left(
Ne^{kj}E_{k}\right)  _{,j}%
\]

\begin{equation}
+\frac{1}{N}\left(  g_{kn}N_{,m}^{n}+g_{mn}N_{,k}^{n}-g_{km,0}+N^{n}%
g_{km,n}\right)  E_{ADM}^{km}\equiv0, \label{eqn46}%
\end{equation}

\begin{equation}
I_{k}^{ADM}=-N_{,k}E-N_{,k}^{m}E_{m}-\left(  N^{m}E_{k}\right)  _{,m}%
+E_{k,0}+\left(  g_{kn}E_{ADM}^{nm}\right)  _{,m}+\left(  g_{km}E_{ADM}%
^{nm}\right)  _{,n}-g_{nm,k}E_{ADM}^{nm}\equiv0. \label{eqn47}%
\end{equation}

We invite the reader to compare (\ref{eqn46})-(\ref{eqn47}) with the covariant
DI of the EH action:%

\begin{equation}
I^{\mu}=\nabla_{\nu}E^{\mu\nu}\equiv0 \label{eqn48}%
\end{equation}
(where $\nabla_{\nu}$ is a covariant derivative)\footnote{The first time this
identity appeared along with the EH action was in Hilbert's work
\cite{Hilbert, Hilbert-eng}.}. We see that the expressions for DIs
(\ref{eqn46})-(\ref{eqn47}) are much larger than (\ref{eqn48}), and the kind
of \textquotedblleft geometrical interpretation\textquotedblright\ of
(\ref{eqn46})-(\ref{eqn47}) that might compel someone to use these DIs and the
ADM variables at the Lagrangian level, is inconceivable. If one were to
consider the Lagrangian method as an algorithm (similar to the Hamiltonian
analysis) to find \textit{a priori} unknown symmetries, a prescription for the
construction of a DI would need to be developed. For covariant theories, it is
natural to expect covariant DIs to exist (e.g. see \cite{Trans}), but it is
not at all clear how to find DIs like (\ref{eqn46})-(\ref{eqn47}). These DIs
can only be explicitly checked by substitution of the ELDs of the ADM action
into (\ref{eqn46})-(\ref{eqn47}).

Let us, as in \cite{KKK-1}, find the group property of transformations
(\ref{eqn40})-(\ref{eqn42}); this entails a more complicated calculation
compared with similar calculations for the EH action. The possibility to work
with a quasi-covariant form leads to considerable simplification (see
\cite{KKK-1}). In this case, separate calculations for the transformations of
different fields are needed. We start from the commutator of two
transformations, and begin with the first field (lapse) -- the one that has
the simplest expression.

We try to present the commutator%

\begin{equation}
\left[  \delta_{2},\delta_{1}\right]  N=\left(  \delta_{2}\delta_{1}%
-\delta_{1}\delta_{2}\right)  N=\delta_{\left[  1,2\right]  }N \label{eqn50}%
\end{equation}
in a form that preserves (\ref{eqn40})%

\begin{equation}
\delta_{\left[  1,2\right]  }N=\varepsilon_{\left[  1,2\right]  ,j}^{\bot
}N^{j}-\varepsilon_{\left[  1,2\right]  ,0}^{\bot}-\varepsilon_{\left[
1,2\right]  }^{i}N_{,i} \label{eqn51}%
\end{equation}
for some expressions, $\varepsilon_{\left[  1,2\right]  }^{\bot}$ and
$\varepsilon_{\left[  1,2\right]  }^{i}$ (to shorten notation, we eliminate
the subscript `$ADM$'). Calculating (\ref{eqn50})%

\[
\left[  \delta_{2},\delta_{1}\right]  N=\delta_{2}\left[  \varepsilon
_{1,j}^{\bot}N^{j}-\varepsilon_{1,0}^{\bot}-\varepsilon_{1}^{i}N_{,i}\right]
-\delta_{1}\left[  \varepsilon_{2,j}^{\bot}N^{j}-\varepsilon_{2,0}^{\bot
}-\varepsilon_{2}^{i}N_{,i}\right]  =
\]

\begin{equation}
\varepsilon_{1,j}^{\bot}\delta_{2}N^{j}-\varepsilon_{1}^{i}\left(  \delta
_{2}N\right)  _{,i}-\varepsilon_{2,j}^{\bot}\delta_{1}N^{j}+\varepsilon
_{2}^{i}\left(  \delta_{1}N\right)  _{,i} \label{eqn52}%
\end{equation}
and using the transformations of fields (\ref{eqn40})-(\ref{eqn41}) and the
field-independence of the parameters (not affected by transformations) gives%

\begin{align}
\left[  \delta_{2},\delta_{1}\right]  N  &  =-\varepsilon_{1,k}^{\bot}\left[
\varepsilon_{2}^{\bot}N_{,j}e^{jk}-\varepsilon_{2,j}^{\bot}Ne^{k^{j}%
}+\varepsilon_{2}^{j}N_{,j}^{k}-\varepsilon_{2,j}^{k}N^{j}+\varepsilon
_{2,0}^{k}\right]  -\varepsilon_{1}^{i}\left[  \varepsilon_{2,k}^{\bot}%
N^{k}-\varepsilon_{2,0}^{\bot}-\varepsilon_{2}^{k}N_{,k}\right]
_{,i}~\nonumber\\
&  +\varepsilon_{2,k}^{\bot}\left[  \varepsilon_{1}^{\bot}N_{,j}%
e^{jk}-\varepsilon_{1,j}^{\bot}Ne^{k^{j}}+\varepsilon_{1}^{j}N_{,j}%
^{k}-\varepsilon_{1,j}^{k}N^{j}+\varepsilon_{1,0}^{k}\right]  +\varepsilon
_{2}^{i}\left[  \varepsilon_{1,k}^{\bot}N^{k}-\varepsilon_{1,0}^{\bot
}-\varepsilon_{1}^{k}N_{,k}\right]  _{,i}~. \label{eqn53}%
\end{align}
Collecting terms with parameters only%

\begin{equation}
-\varepsilon_{1,k}^{\bot}\varepsilon_{2,0}^{k}+\varepsilon_{1}^{k}%
\varepsilon_{2,0k}^{\bot}+\varepsilon_{2,k}^{\bot}\varepsilon_{1,0}%
^{k}-\varepsilon_{2}^{k}\varepsilon_{1,0k}^{\bot}=-\left(  -\varepsilon
_{1}^{k}\varepsilon_{2,k}^{\bot}+\varepsilon_{2}^{k}\varepsilon_{1,k}^{\bot
}\right)  _{,0} \label{eqn54}%
\end{equation}
and comparing with (\ref{eqn51}) (the term without fields is $-\varepsilon
_{\left[  1,2\right]  ,0}^{\bot}$), we uniquely obtain%

\begin{equation}
\varepsilon_{\left[  1,2\right]  }^{\perp}=-\varepsilon_{1}^{k}\varepsilon
_{2,k}^{\bot}+\varepsilon_{2}^{k}\varepsilon_{1,k}^{\bot}~. \label{eqn55}%
\end{equation}

The remaining terms (see (\ref{eqn53})) must be combined into $\varepsilon
_{\left[  1,2\right]  ,j}^{\bot}N^{j}$ with the same $\varepsilon_{\left[
1,2\right]  }^{\perp}$; what is left has to contribute to the second parameter%

\begin{equation}
\varepsilon_{\left[  1,2\right]  }^{k}=\varepsilon_{1,m}^{\bot}\varepsilon
_{2}^{\bot}e^{mk}-\varepsilon_{1}^{i}\varepsilon_{2,i}^{k}-\varepsilon
_{2,m}^{\bot}\varepsilon_{1}^{\bot}e^{mk}+\varepsilon_{2}^{i}\varepsilon
_{1,i}^{k}~. \label{eqn56}%
\end{equation}
Note that the same redefinition of parameters, as was made in the EH
Lagrangian (Eq. (37) of \cite{KKK-1}), is responsible for a breakdown of the
group properties of the ADM transformations.

We must check the consistency of the structure of parameters (\ref{eqn55}%
)-(\ref{eqn56}), which are a kind of \textquotedblleft structure
functions\textquotedblright,\ for the rest of the fields. We know what kind of
the parameter to expect, and this simplifies the calculations (gives some hint
how to sort out terms). A straightforward calculation leads to the result that
the same redefinition works for the shift and spatial components of the metric
tensor, i.e. it is consistent for all fields:%

\begin{equation}
\left[  \delta_{2},\delta_{1}\right]  \left(
\begin{array}
[c]{c}%
N\\
N^{i}\\
g_{km}%
\end{array}
\right)  =\delta_{\left[  1,2\right]  }\left(
\begin{array}
[c]{c}%
N\\
N^{i}\\
g_{km}%
\end{array}
\right)  \label{eqn57}%
\end{equation}
with $\varepsilon_{\left[  1,2\right]  }^{\perp}$ and $\varepsilon_{\left[
1,2\right]  }^{k}$ given by (\ref{eqn55})-(\ref{eqn56}). Note that the
explicit form of the transformations is different for the fields in
(\ref{eqn57}), but the composition of the parameters is the same.

To determine whether or not these transformations form a group, the double
commutator is needed to check the Jacobi identity:%

\begin{equation}
\left(  \left[  \left[  \delta_{1},\delta_{2}\right]  ,\delta_{3}\right]
+\left[  \left[  \delta_{3},\delta_{1}\right]  ,\delta_{2}\right]  +\left[
\left[  \delta_{2},\delta_{3}\right]  ,\delta_{1}\right]  \right)  \left(
\begin{array}
[c]{c}%
N\\
N^{i}\\
g_{km}%
\end{array}
\right)  \equiv0. \label{eqn60}%
\end{equation}
In this case, if it is not zero for one field, then (\ref{eqn60}) is not an
identity and the transformations do not form a group.

Let us consider the double commutator for the field with the simplest transformation%

\begin{equation}
\left[  \left[  \delta_{1},\delta_{2}\right]  ,\delta_{3}\right]  N=\left(
\delta_{\left[  1,2\right]  }\delta_{3}-\delta_{3}\delta_{\left[  1,2\right]
}\right)  N~; \label{eqn61}%
\end{equation}
in a manner similar to (\ref{eqn52}), we obtain%

\[
\delta_{\left[  1,2\right]  }\left[  \varepsilon_{3,j}^{\bot}N^{j}%
-\varepsilon_{3,0}^{\bot}-\varepsilon_{3}^{i}N_{,i}\right]  -\delta_{3}\left[
\varepsilon_{\left[  1,2\right]  ,j}^{\bot}N^{j}-\varepsilon_{\left[
1,2\right]  ,0}^{\bot}-\varepsilon_{\left[  1,2\right]  }^{i}N_{,i}\right]
\]

\begin{equation}
=\varepsilon_{3,j}^{\bot}\delta_{\left[  1,2\right]  }N^{j}-\varepsilon
_{3}^{i}\left(  \delta_{\left[  1,2\right]  }N\right)  _{,i}-\varepsilon
_{\left[  1,2\right]  ,j}^{\bot}\delta_{3}N^{j}+\varepsilon_{\left[
1,2\right]  }^{i}\left(  \delta_{3}N\right)  _{,i}+N_{,i}\delta_{3}%
\varepsilon_{\left[  1,2\right]  }^{i}~. \label{eqn62}%
\end{equation}

The first four terms on the right hand side of (\ref{eqn62}) are the same as
those in (\ref{eqn52}), and only the last one (the fifth term)\ creates the
additional contribution. Consequently, by making a simple substitution of
indices in (\ref{eqn55})-(\ref{eqn56}): $1\rightarrow\left[  1,2\right]  $ and
$2\rightarrow3$ (as was performed in \cite{KKK-1}), the first four terms lead
to the same results for the parameters. Due to the last contribution in
(\ref{eqn62}), one extra term will appear, for the $\varepsilon_{\left[
\left[  1,2\right]  ,3\right]  }^{k}$ parameter,%

\begin{equation}
\varepsilon_{\left[  \left[  1,2\right]  ,3\right]  }^{\perp}=-\varepsilon
_{\left[  1,2\right]  }^{k}\varepsilon_{3,k}^{\bot}+\varepsilon_{3}%
^{k}\varepsilon_{\left[  1,2\right]  ,k}^{\bot}~, \label{eqn63}%
\end{equation}

\begin{equation}
\varepsilon_{\left[  \left[  1,2\right]  ,3\right]  }^{k}=\varepsilon_{\left[
1,2\right]  ,m}^{\bot}\varepsilon_{3}^{\bot}e^{mk}-\varepsilon_{\left[
1,2\right]  }^{i}\varepsilon_{3,i}^{k}-\varepsilon_{3,m}^{\bot}\varepsilon
_{\left[  1,2\right]  }^{\bot}e^{mk}+\varepsilon_{3}^{i}\varepsilon_{\left[
1,2\right]  ,i}^{k}+\delta_{3}\varepsilon_{\left[  1,2\right]  }^{k}~.
\label{eqn64}%
\end{equation}

To obtain an explicit form, the expressions for $\varepsilon_{\left[
1,2\right]  }^{\bot}$ and $\varepsilon_{\left[  1,2\right]  }^{k}$,
(\ref{eqn55}) and (\ref{eqn56}), must be substituted into (\ref{eqn64}); in
the last term only the part of $\varepsilon_{\left[  1,2\right]  }^{k}$ that
is proportional to fields is needed since only fields are affected by
$\delta_{3}$. The final expression is%

\begin{equation}
\delta_{3}\varepsilon_{\left[  1,2\right]  }^{k}=\left[  \varepsilon
_{1,m}^{\bot}\varepsilon_{2}^{\bot}-\varepsilon_{2,m}^{\bot}\varepsilon
_{1}^{\bot}\right]  \delta_{3}e^{mk}=-\left[  \varepsilon_{1,m}^{\bot
}\varepsilon_{2}^{\bot}-\varepsilon_{2,m}^{\bot}\varepsilon_{1}^{\bot}\right]
e^{kp}e^{mn}\delta_{3}g_{pn} \label{eqn65}%
\end{equation}

\[
=\left(  \varepsilon_{2,m}^{\bot}\varepsilon_{1}^{\bot}-\varepsilon
_{1,m}^{\bot}\varepsilon_{2}^{\bot}\right)  \left[  \varepsilon_{3}^{\bot
}\frac{1}{N}\left(  e^{mn}N_{,n}^{k}+e^{kp}N_{,p}^{m}+e_{,0}^{mk}-e_{,i}%
^{km}N^{i}\right)  -\varepsilon_{3,p}^{m}e^{kp}-\varepsilon_{3,n}^{k}%
e^{mn}+\varepsilon_{3}^{i}e_{,i}^{km}\right]  ,
\]
where $g_{km}e^{mn}=\delta_{k}^{n}$ was used to find $\delta_{3}e^{mk}$ from
(\ref{eqn42}).

Verifying the Jacobi identity (\ref{eqn60}) is equivalent to checking the
corresponding identity for the parameters, which for $\varepsilon^{k}$ is:%

\begin{equation}
\varepsilon_{\left[  \left[  1,2\right]  ,3\right]  }^{k}+\varepsilon_{\left[
\left[  3,1\right]  ,2\right]  }^{k}+\varepsilon_{\left[  \left[  2,3\right]
,1\right]  }^{k}\neq0. \label{eqn66}%
\end{equation}

The transformations that follow from the Hamiltonian formulation of the ADM
action do not form a group; this is exactly the same outcome as for the case
of the EH action, where the ADM symmetry was one of many symmetries that can
be found at the Lagrangian level by a non-covariant modification of the DI for
the diffeomorphism transformation \cite{KKK-1}. The only difference is that
for the EH action, the Hamiltonian formulation leads to diffeomorphism
invariance instead of the ADM transformations for the Hamiltonian formulation
of the ADM action. This difference in symmetries produced by the Dirac
procedure is naturally related to the non-canonicity of these variables
between the two Hamiltonians \cite{Myths}. To complete the comparison, we have
to examine the properties of both symmetries, as we did for the EH action
\cite{KKK-1}. Let us find the diffeomorphism transformation for the ADM
variables, and check whether or not it is also a symmetry of the ADM action.
Note that we will consider the full diffeomorphism, not only its
\textquotedblleft spatial\textquotedblright\ part, the part that some
consider\ to be a symmetry of the ADM formulation, and consequently a symmetry
of GR (e.g., see \cite{Pullin-18}: \textquotedblleft Unfortunately, the
canonical treatment breaks the symmetry between space and time in general
relativity and the resulting algebra of constraints is not the algebra of four
diffeomorphisms\textquotedblright).

\section{Is diffeomorphism a gauge symmetry of the ADM action?}

Let us first determine how the ADM variables should transform under
diffeomorphism. The known transformations of the contravariant metric,%

\begin{equation}
\delta_{\mathit{diff}}g^{\mu\nu}=\xi_{,\alpha}^{\nu}g^{\mu\alpha}+\xi
_{,\alpha}^{\mu}g^{\nu\alpha}-g_{,\alpha}^{\mu\nu}\xi^{\alpha}, \label{eqn70}%
\end{equation}
and the covariant metric with the same, contravariant gauge parameter
$\xi^{\alpha}$,%

\begin{equation}
\delta_{\mathit{diff}}g_{\gamma\sigma}=-g_{\sigma\nu}\xi_{,\gamma}^{\nu
}-g_{\gamma\mu}\xi_{,\sigma}^{\mu}-g_{\gamma\sigma,\alpha}\xi^{\alpha},
\label{eqn71}%
\end{equation}
are needed because the ADM variables are a mixture of covariant and
contravariant components of the metric tensor (\ref{eqn10}). (Equation
(\ref{eqn71}) can be obtained from the condition $g_{\mu\nu}g^{\nu\alpha
}=\delta_{\mu}^{\alpha}$.)

Using change of variables (\ref{eqn10}), one finds the diffeomorphism
transformations of the ADM variables in terms of the metric by applying
(\ref{eqn70})%

\begin{equation}
\delta_{\mathit{diff}}N=\delta_{\mathit{diff}}\left(  -g^{00}\right)
^{-1/2}=-\frac{1}{2}\left(  -g^{00}\right)  ^{-3/2}\left(  -\delta
_{\mathit{diff}}g^{00}\right)  . \label{eqn73}%
\end{equation}
Substituting $\delta_{\mathit{diff}}g^{00}$ from (\ref{eqn70}) and expressing
the metric components in (\ref{eqn73}) in terms of the ADM variables, one
obtains for the lapse $N$:%

\begin{equation}
\delta_{\mathit{diff}}N=-\left(  N\xi^{0}\right)  _{,0}+N\xi_{,k}^{0}N^{k}%
-\xi^{k}N_{,k}~. \label{eqn74}%
\end{equation}
Repeating the same calculations for the shift $N^{k}$ and space-space
components $g_{km}$ yields:%

\begin{equation}
\delta_{\mathit{diff}}N^{k}=+\xi_{,m}^{0}N^{2}e^{im}-\xi_{,0}^{i}+\xi_{,p}%
^{i}N^{p}-N_{,0}^{i}\xi^{0}-N_{,k}^{i}\xi^{k}-N^{i}\xi_{,0}^{0}+N^{i}\xi
_{,k}^{0}N^{k}, \label{eqn75}%
\end{equation}

\begin{equation}
\delta_{\mathit{diff}}g_{km}=-g_{kp}N^{p}\xi_{,m}^{0}-g_{kp}\xi_{,m}%
^{p}-g_{mp}N^{p}\xi_{,k}^{0}-g_{mp}\xi_{,k}^{p}-g_{km,0}\xi^{0}-g_{km,p}%
\xi^{p}. \label{eqn76}%
\end{equation}

If these transformations are a gauge symmetry of the ADM action, then the
corresponding Noether's DIs must exist and they can be easily restored,
similarly to (\ref{eqn45}) of the previous Section,%

\begin{equation}
\delta S_{ADM}=\int\left[  E\delta_{\mathit{diff}}N+E_{i}\delta_{\mathit{diff}%
}N^{i}+E_{ADM}^{km}\delta_{\mathit{diff}}g_{km}\right]  d^{4}x=\int\left[
\xi^{0}I_{0}^{\mathit{diff}}+\xi^{k}I_{k}^{\mathit{diff}}\right]  d^{4}x.
\label{eqn79}%
\end{equation}
Simple rearrangements then lead to:%

\[
I_{0}^{\mathit{diff}}=+NE_{,0}-\left(  NN^{k}E\right)  _{,k}-\left(
N^{2}e^{im}E_{i}\right)  _{,m}-N_{,0}^{i}E_{i}+\left(  N^{i}E_{i}\right)
_{,0}-\left(  N^{i}N^{k}E_{i}\right)  _{,k}%
\]

\begin{equation}
+\left(  g_{kp}N^{p}E_{ADM}^{km}\right)  _{,m}+\left(  g_{mp}N^{p}E_{ADM}%
^{km}\right)  _{,k}-g_{km,0}E_{ADM}^{km}\equiv0, \label{eqn80}%
\end{equation}

\begin{equation}
I_{k}^{\mathit{diff}}=-N_{,k}E+E_{k,0}-\left(  N^{p}E_{k}\right)  _{,p}%
-N_{,k}^{i}E_{i}+\left(  g_{nk}E_{ADM}^{nm}\right)  _{,m}+\left(
g_{mk}E_{ADM}^{nm}\right)  _{,n}-g_{nm,k}E_{ADM}^{nm}\equiv0. \label{eqn81}%
\end{equation}

The correctness of these DIs can be checked by direct substitution of the ELDs
of the ADM action into (\ref{eqn80})-(\ref{eqn81}). Because the DIs for ADM
transformations are already known, (\ref{eqn46})-(\ref{eqn47}), to prove that
(\ref{eqn80})-(\ref{eqn81}) are indeed DIs (and according to the converse of
Noether's second theorem, the corresponding transformations are a gauge
invariance of the ADM action), it is enough to show that they are a linear
combination of the known DIs. Let us seek such connections. Compare DI
(\ref{eqn47}) for the ADM transformation and DI (\ref{eqn81}) for
diffeomorphism; we find that they are identical,%

\begin{equation}
I_{k}^{\mathit{diff}}=I_{k}^{ADM}. \label{eqn82}%
\end{equation}

DI (\ref{eqn46}) for the ADM transformations and DI (\ref{eqn80}) for
diffeomorphism are not the same; but we observe that the terms with time
derivatives of the ELDs for the lapse in (\ref{eqn46}) is $+E_{,0}~$, and in
(\ref{eqn80}) it is $+NE_{,0}~$. This suggests the relation $I_{0}%
^{\mathit{diff}}=NI_{\bot}^{ADM}+...$. After collecting such terms, the
analysis of the remainder leads to the extra contribution proportional to
$I_{k}^{ADM}$,%

\begin{equation}
I_{0}^{\mathit{diff}}=NI_{\bot}^{ADM}+N^{k}I_{k}^{ADM}. \label{eqn83}%
\end{equation}

Because the DIs of diffeomorphism are linear combinations of DIs known for the
ADM transformations, diffeomorphism is also a symmetry of the ADM action. Let
us perform some simple rearrangements,%

\[
\xi^{k}I_{k}^{\mathit{diff}}+\xi^{0}I_{0}^{\mathit{diff}}=\xi^{k}I_{k}%
^{ADM}+\xi^{0}\left(  NI_{\bot}^{ADM}+N^{k}I_{k}^{ADM}\right)
\]

\begin{equation}
=\xi^{0}NI_{\bot}^{ADM}+\left(  \xi^{k}+\xi^{0}N^{k}\right)  I_{k}%
^{ADM}=\tilde{\varepsilon}^{\bot}I_{\bot}^{ADM}+\tilde{\varepsilon}^{k}%
I_{k}^{ADM}; \label{eqn84}%
\end{equation}
this allows us to present the diffeomorphism transformations of the ADM action
in a particular form, in which the ADM transformations have field-dependent parameters:%

\begin{equation}
\tilde{\varepsilon}^{\bot}=N\xi^{0}, \label{eqn85}%
\end{equation}

\begin{equation}
\tilde{\varepsilon}^{k}=\xi^{0}N^{k}+\xi^{k}. \label{eqn86}%
\end{equation}

Note that all of the different symmetries, which can be constructed using
linear combinations of DIs, can be presented in such a form where a
field-dependent redefinition of parameters appears. This fact does not make
these DIs equal because the gauge parameters must be independent of fields;
consequently, such relations (such as (\ref{eqn85}) and (\ref{eqn86})) can be
used only as shorthand notation that might simplify some calculations. In the
case of the EH action, it was possible to use such a presentation to write
different transformations in quasi-covariant form, and thus perform the
calculations at once for all of the components of the metric tensor
\cite{KKK-1}. For the ADM formulation, this technique is not possible because
of its non-covariant form; but relations (\ref{eqn85})-(\ref{eqn86}) allow us
to use the results of the previous Section to help in the calculation of the
commutators of the two transformations for the ADM variables (we are
interested in group properties of transformations (\ref{eqn74})-(\ref{eqn76})).

The transformations under diffeomorphism can be presented in the form of ADM
transformations with field-dependent parameters (which are shorthand notations
(\ref{eqn85})-(\ref{eqn86}))%

\begin{equation}
\delta_{\mathit{diff}}N=\tilde{\varepsilon}_{,k}^{\bot}N^{k}-\tilde
{\varepsilon}_{,0}^{\bot}-\tilde{\varepsilon}^{k}N_{,k}~; \label{eqn90}%
\end{equation}
and we can find the commutator (to shorten notation, we use $\delta
_{\mathit{diff}}\equiv\tilde{\delta}$)%

\[
\left(  \tilde{\delta}_{2}\tilde{\delta}_{1}-\tilde{\delta}_{1}\tilde{\delta
}_{2}\right)  N=\tilde{\delta}_{2}\left(  \tilde{\varepsilon}_{1,k}^{\bot
}N^{k}-\tilde{\varepsilon}_{1,0}^{\bot}-\tilde{\varepsilon}_{1}^{k}%
N_{,k}\right)  -\tilde{\delta}_{1}\left(  \tilde{\varepsilon}_{2,k}^{\bot
}N^{k}-\tilde{\varepsilon}_{2,0}^{\bot}-\tilde{\varepsilon}_{2}^{k}%
N_{,k}\right)
\]
that, taking into account the field-dependent representation of parameters
(\ref{eqn85})-(\ref{eqn86}), can be written as%

\begin{equation}
\left(  \tilde{\delta}_{2}\tilde{\delta}_{1}-\tilde{\delta}_{1}\tilde{\delta
}_{2}\right)  N=\tilde{\varepsilon}_{1,k}^{\bot}\tilde{\delta}_{2}N^{k}%
-\tilde{\varepsilon}_{1}^{k}\tilde{\delta}_{2}N_{,k}-\tilde{\varepsilon}%
_{2,k}^{\bot}\tilde{\delta}_{1}N^{k}+\tilde{\varepsilon}_{2}^{k}\tilde{\delta
}_{1}N_{,k}\text{ \ \ } \label{eqn94}%
\end{equation}

\[
+N^{k}\tilde{\delta}_{2}\tilde{\varepsilon}_{1,k}^{\bot}-\tilde{\delta}%
_{2}\tilde{\varepsilon}_{1,0}^{\bot}-N_{,k}\tilde{\delta}_{2}\tilde
{\varepsilon}_{1}^{k}-N^{k}\tilde{\delta}_{1}\tilde{\varepsilon}_{2,k}^{\bot
}+\tilde{\delta}_{1}\tilde{\varepsilon}_{2,0}^{\bot}+N_{,k}\tilde{\delta}%
_{1}\tilde{\varepsilon}_{2}^{k}~.
\]

The first line corresponds exactly to the previous calculations (\ref{eqn52})
(of course, with a new field-dependent parameter), so we can use the previous
result, and substitute parameters (\ref{eqn85})-(\ref{eqn86}) into
(\ref{eqn94}). The last line of (\ref{eqn94}) gives additional terms, which
can be combined to obtain%

\[
N^{k}\left(  \tilde{\delta}_{2}\tilde{\varepsilon}_{1}^{\bot}-\tilde{\delta
}_{1}\tilde{\varepsilon}_{2}^{\bot}\right)  _{,k}-\left(  \tilde{\delta}%
_{2}\tilde{\varepsilon}_{1}^{\bot}-\tilde{\delta}_{1}\tilde{\varepsilon}%
_{2}^{\bot}\right)  _{,0}-\left(  \tilde{\delta}_{2}\tilde{\varepsilon}%
_{1}^{k}-\tilde{\delta}_{1}\tilde{\varepsilon}_{2}^{k}\right)  N_{,k}~.
\]
The composition of the parameters is now given by%

\[
\tilde{\varepsilon}_{\left[  1,2\right]  }^{\bot}=-\tilde{\varepsilon}_{1}%
^{k}\tilde{\varepsilon}_{2,k}^{\bot}+\tilde{\varepsilon}_{2}^{k}%
\tilde{\varepsilon}_{1,k}^{\bot}+\tilde{\delta}_{2}\tilde{\varepsilon}%
_{1}^{\bot}-\tilde{\delta}_{1}\tilde{\varepsilon}_{2}^{\bot}%
\]

\begin{equation}
=-\tilde{\varepsilon}_{1}^{k}\tilde{\varepsilon}_{2,k}^{\bot}+\tilde
{\varepsilon}_{2}^{k}\tilde{\varepsilon}_{1,k}^{\bot}+\xi_{1}^{0}\tilde
{\delta}_{2}N-\xi_{2}^{0}\tilde{\delta}_{1}N~. \label{eqn96}%
\end{equation}
Performing further calculation gives%

\[
\tilde{\varepsilon}_{\left[  1,2\right]  }^{\bot}=N\left(  -\xi_{1}^{k}%
\xi_{2,k}^{0}+\xi_{2}^{k}\xi_{1,k}^{0}-\xi_{1}^{0}\xi_{2,0}^{0}+\xi_{2}^{0}%
\xi_{1,0}^{0}\right)  =N\xi_{\left[  1,2\right]  }^{0}=N\left(  -\xi
_{1}^{\alpha}\xi_{2,\alpha}^{0}+\xi_{2}^{\alpha}\xi_{1,\alpha}^{0}\right)  ,
\]
with%

\begin{equation}
\xi_{\left[  1,2\right]  }^{0}=\xi_{2}^{\alpha}\xi_{1,\alpha}^{0}-\xi
_{1}^{\alpha}\xi_{2,\alpha}^{0}~. \label{eqn99}%
\end{equation}
Note that a covariant form is partially restored. The same can be confirmed
for the second parameter,%

\begin{equation}
\tilde{\varepsilon}_{\left[  1,2\right]  }^{k}=\varepsilon_{,m}^{1\bot
}\varepsilon^{2\bot}e^{mk}-\varepsilon^{1i}\varepsilon_{,i}^{2k}%
-\varepsilon_{,m}^{2\bot}\varepsilon^{1\bot}e^{mk}+\varepsilon^{2i}%
\varepsilon_{,i}^{1k}+\tilde{\delta}_{2}\tilde{\varepsilon}_{1}^{k}%
-\tilde{\delta}_{1}\tilde{\varepsilon}_{2}^{k}~, \label{eqn101}%
\end{equation}
which can also be presented in semi-covariant form%

\begin{equation}
\tilde{\varepsilon}_{\left[  1,2\right]  }^{k}=\xi_{\left[  1,2\right]  }%
^{k}=\xi_{2}^{\alpha}\xi_{1,\alpha}^{k}-\xi_{1}^{\alpha}\xi_{2,\alpha}^{k}~.
\label{eqn102}%
\end{equation}

In spite of using non-covariant ADM variables for the diffeomorphism
transformations, which are not only complicated expressions, but different for
each field ($N,N^{k},g_{km}$), the parameters are redefined in a covariant
way. The combination of (\ref{eqn99}) and (\ref{eqn102}) obviously can be
written in a covariant form%

\begin{equation}
\xi_{\left[  1,2\right]  }^{\mu}=\xi_{2}^{\alpha}\xi_{1,\alpha}^{\mu}-\xi
_{1}^{\alpha}\xi_{2,\alpha}^{\mu}~. \label{eqn104}%
\end{equation}

Because the variables are not covariant, the complete proof is more
complicated. To verify (\ref{eqn104}) for all fields similar calculations must
be repeated for the remaining fields, and the consistency of the redefinition
of parameters must be checked. Such calculations confirm that the above
redefinition is preserved for all fields, i.e.%

\begin{equation}
\left(  \tilde{\delta}_{2}\tilde{\delta}_{1}-\tilde{\delta}_{1}\tilde{\delta
}_{2}\right)  \left(
\begin{array}
[c]{c}%
N\\
N^{i}\\
g_{km}%
\end{array}
\right)  =\tilde{\delta}_{\left[  1,2\right]  }\left(
\begin{array}
[c]{c}%
N\\
N^{i}\\
g_{km}%
\end{array}
\right)  . \label{eqn105}%
\end{equation}

Fields are absent in (\ref{eqn99}) and (\ref{eqn102}), that is why the
correctness of the Jacobi identity for double commutators is evident, i.e. the
group properties are preserved. So among the four cases (a) - (d), which are
discussed in \cite{KKK-1}, the last one, (d), is realized -- the EH and ADM
actions are invariant under diffeomorphism, the only symmetry with the group
property. But this result raised many questions, some of which were briefly
discussed in \cite{KKK-1}.

One question is: what is the origin of the statements, so often made in the
literature, that the ADM action is invariant only under spatial
diffeomorphism? As we have shown in the present paper, the ADM variables (all
of them: $N,N^{i}$ and $g_{km}$) are invariant under the diffeomorphism
transformations (\ref{eqn74})-(\ref{eqn76}) that correspond to the DIs
(\ref{eqn82})-(\ref{eqn83}). Although these DIs are non-covariant, they lead
to the invariance of the ADM action under 4-diffeomorphism. Therefore the
assertion of Kuchar and Isham\ \cite{Isham} that \textquotedblleft the full
group of spacetime diffeomorphism has somehow got lost in making the
transition from the Hilbert action to the Dirac-ADM action\textquotedblright%
\ is not correct. The statements about spatial diffeomorphism can even be
found in earlier papers that describe the first attempts to perform the
Hamiltonian analysis of GR. For example, in \cite{Higgs} it is written:
\textquotedblleft It is clear that $\mathcal{H}_{u}\left(  x\right)  $
[\footnote{In \textquotedblleft modern\textquotedblright\ language:
\textquotedblleft diffeomorphism\textquotedblright, or \textquotedblleft
momentum\textquotedblright,\ constraint.}] is just the set of infinitesimal
generators of the group of general coordinate transformations on the potential
$g_{rs}$\textquotedblright\footnote{In Erratum \cite{Erratum}, Higgs commented
that his \textquotedblleft former statement is not quite
correct\textquotedblright\ and \textquotedblleft certain transversality
conditions\textquotedblright\ must be satisfied.} (see also \cite{DeWitt}) and
continue to propagate to more recent papers, e.g. in \cite{Pullin-18}:
\textquotedblleft the diffeomorphism constraint can be shown to be associated
with the invariance of general relativity under spatial
diffeomorphism\textquotedblright, or in \cite{Shestakova2004}:
\textquotedblleft the momentum constraints ... generate diffeomorphism of
3-metric $g_{ab}$\textquotedblright.

As we have shown in \cite{Myths}, a spatial diffeomorphism alone cannot be
obtained directly in the course of the Hamiltonian analysis of the ADM
formulation without some unjustified manipulations, such as: disregarding
primary first-class constraints\footnote{To quote Dirac \cite{Diracbook}%
:\textquotedblleft If we are to have any motion at all with a zero
Hamiltonian, we must have at least one primary constraint.\textquotedblright};
promoting secondary constraints into primary; and leaving only one,
\textquotedblleft diffeomorphism\textquotedblright, constraint in the gauge
generator, which would produce spatial diffeomorphism only for $g_{km}$. But
such manipulations contradict any procedure for finding gauge transformations
and cannot be seriously considered. The Dirac Hamiltonian procedure applied to
the ADM action leads to different gauge transformations: (\ref{eqn40}%
)-(\ref{eqn42}). To derive the gauge transformations for the ADM or EH
Hamiltonian formulations, all the first-class constraints, four primary and
four secondary, are needed. Using only three constraints to produce a splinter
of gauge transformations is not understandable, and a \textquotedblleft
geometrical interpretation\textquotedblright\ is no justification.

The same is valid at the Lagrangian level. According to Noether's theorem,
there are four independent DIs for the ADM or EH action, therefore, there are
four independent parameters in the gauge transformations. Let us assume that
someone wants to use only three DIs, for example $I_{k}\equiv0$, and to set
$\tilde{\varepsilon}^{\bot}=0$, then the commutator of such transformations
gives for both (\ref{eqn40})-(\ref{eqn42}) and (\ref{eqn74})-(\ref{eqn76})%

\begin{equation}
\left(  \tilde{\delta}_{\left(  k\right)  2}\tilde{\delta}_{\left(  k\right)
1}-\tilde{\delta}_{\left(  k\right)  1}\tilde{\delta}_{\left(  k\right)
2}\right)  \left(
\begin{array}
[c]{c}%
N\\
N^{i}\\
g_{km}%
\end{array}
\right)  =\tilde{\delta}_{\left(  k\right)  \left[  1,2\right]  }\left(
\begin{array}
[c]{c}%
N\\
N^{i}\\
g_{km}%
\end{array}
\right)  \label{eqn106}%
\end{equation}
with%

\[
\xi_{\left[  1,2\right]  }^{k}=\xi_{2}^{m}\xi_{1,m}^{k}-\xi_{1}^{m}\xi
_{2,m}^{k}%
\]
(the subscript $\left(  k\right)  $ in $\tilde{\delta}_{\left(  k\right)  i}$
indicates that these transformations correspond to a spatial DI $I_{k}\equiv0$ only).

From (\ref{eqn106}) it looks as though spatial transformations can be
considered separately from the translation in the time direction. The
situation is even more confusing because of (\ref{eqn82}), which one might
take as an indication of the presence of spatial diffeomorphism. Firstly, note
that all fields are engaged in (\ref{eqn106}), not just the spatial components
of the metric tensor. Secondly, the commutator of two transformations that
involves only a \textquotedblleft perpendicular\textquotedblright\ parameter
$\tilde{\varepsilon}_{i}^{\bot}$ cannot be written in the form of
\textquotedblleft perpendicular\textquotedblright\ transformation with the new
parameter $\tilde{\varepsilon}_{\left[  1,2\right]  }^{\bot}$, as was
similarly done in (\ref{eqn96}) (note that $\tilde{\varepsilon}_{\left[
1,2\right]  }^{\bot}$ in (\ref{eqn96}) also depends on $\tilde{\varepsilon
}_{i}^{k}$, so the \textquotedblleft perpendicular\textquotedblright%
\ transformations do not form a group). And thirdly, DIs (\ref{eqn80}) and
(\ref{eqn81}) are satisfied identically if \textit{all} ELDs are present. For
example, if one takes the last three terms in (\ref{eqn80}) or (\ref{eqn81}),
which are the only terms responsible for the spatial diffeomorphism of
$g_{km}$, then such identities would vanish only if $E=0$ and $E_{k}=0$; but
this can only happen if the corresponding fields, $N$ and $N^{i}$, are not
present in the Lagrangian (or $N$ and $N^{i}$ are constants). In such a case,
however, the remainder of the action will not represent a gauge theory at all
because it will be quadratic in the first derivatives in time of $g_{km}$,
which is an invertible expression.

In relation to the Hamiltonian formulation (puzzle), there are other questions
to ask: why is it that when working with the original variable, the metric,
the transformation that follows from the constraint structure of the PSS and
Dirac Hamiltonian formulations is diffeomorphism? This is the only symmetry
with a group property, out of the many transformations that can be constructed
using modifications of the DIs for the EH action \cite{KKK-1}; yet when the
same Dirac method is applied to the ADM action, a different symmetry follows,
and because this symmetry is different, the two formulations are not
canonically related. Is it \textquotedblleft the contradiction that again
witnesses about the incompleteness of theoretical foundation\textquotedblright%
\ \cite{ShestakovaCQG}? Working with the original Einstein's variables, the
Hamiltonian procedure produces diffeomorphism transformations as in all field
theories when the original, natural choice of variables is used; there is
nothing puzzling in such a result. Is it puzzling that the Dirac procedure
gives a different symmetry for the ADM action? \textquotedblleft Would not it
be better to restrict ourselves by transformations in phase space of original
canonical variables in the sense of Dirac?\textquotedblright%
\ \cite{ShestakovaCQG}; this solution has to be rejected according to
\cite{ShestakovaCQG} by the reasoning that the ADM parametrization
\textquotedblleft is preferable because of its geometrical
interpretation\textquotedblright. But what about many other possible changes
of field variables? \ Are they not also subject to this criterion? \ And why
is it that their possible geometrical interpretation also not significant or
preferable? Do we need this plurality\footnote{The explanation of why we have
a plurality of parametrizations, but only one should be chosen, is similar to
the reason why Einstein rejected extra dimensions \cite{Einstein-Science}:
\textquotedblleft It is anomalous to replace the four-dimensional continuum by
a five-dimensional one and then subsequently to tie up artificially one of
these five dimensions in order to account for the fact that it does not
manifest itself\textquotedblright.}?

The Dirac method does not produce diffeomorphism transformations for the ADM
variables (or for any other possible parametrization of fields, except the
original variable -- the metric); and because of this fact, it is claimed in
\cite{ShestakovaCQG} that the Dirac method is \textquotedblleft not
fundamental and undoubted\textquotedblright,\ which suggests that the puzzle
is related to the Dirac method. These suppositions are rooted in the choice of
preferable geometrical interpretation. If someone is satisfied with the
geometrical meaning of the metric and with the original formulation of
Einstein, then no puzzle exists, and the Dirac approach is fundamental; but if
another, geometrical meaning (or interpretation) is preferred\footnote{There
are many possible reasons, e.g.\textquotedblleft\ to recover the old comforts
of a Hamiltonian-like scheme: a system of hypersurfaces stacked in a
well-defined way in space-time, with the system of dynamical variables
distributed over these hypersurfaces and developing uniquely from one
hypersurface to another\textquotedblright\ \cite{Kuchar}, or in \cite{Haw-Pen}%
: \textquotedblleft\ Although `reasonable' from the point of view of classical
Laplacian determinism, the assumption of the existence of a global Cauchy
hypersurface is hard to justify from the standpoint of general
relativity.\textquotedblright}, which is considered more fundamental than the
geometrical meaning of the metric, then the Dirac approach is judged not to be
fundamental and it must be modified or substituted with another method.

Would it not be better to argue that the ADM variables (and all other possible
parametrizations, except the metric) are not fundamental and undoubted,
instead of the Dirac method, and that the geometrical interpretation of the
ADM variables is in contradiction with the geometrical meaning of Einstein's
theory? We provide such arguments in Discussion.

\section{Discussion}

In our previous paper \cite{KKK-1}, we analyzed the group properties of
different symmetries of the EH action. We found that among the plurality of
gauge symmetries and gauge transformations, which can be built by modifying
Noether's DIs of the EH action, there is one symmetry, diffeomorphism, with a
group property \cite{KKK-1}. Other DIs can be obtained by writing different
linear combinations of known DIs. These new DIs correspond to field-dependent
redefinitions of gauge parameters in a formal way. Such a correspondence
should not be taken as another representation of the same symmetry because
these DIs describe different symmetries. This can be said about the ADM
transformations: despite being called the\ \textquotedblleft field-dependent
diffeomorphism\textquotedblright, they differ from diffeomorphism and do not
have group properties \cite{KKK-1}. Further, such modifications as were made
to obtain the ADM transformations destroy the covariant character of the basic
Noether's DI (\ref{eqn48}) of the EH action; and in addition to the
disappearance of group properties, such DIs also effectively impose severe
restrictions on possible coordinate transformations (in generally covariant
theory!): \textquotedblleft the most general set of coordinate transformations
is reduced to arbitrary 3-dimensional transformations and time
reparametrization\textquotedblright\ \cite{CLM}.

In the present paper we performed an analysis similar to that in \cite{KKK-1}
for the ADM action: studying group properties under the same transformations
-- diffeomorphism and the ADM symmetry. As in the case of the EH action, these
two (and many more) symmetries leave the action invariant, but only
diffeomorphism has a group property, despite diffeomorphism transformations
for the ADM action having a very different form compared to those of the EH
action. As we have mentioned before, this result contradicts the statement of
Isham and Kuchar \cite{Isham}: \textquotedblleft the full group of spacetime
diffeomorphisms has somehow got lost in making the transition from the Hilbert
action to the Dirac-ADM action\textquotedblright. In general, any symmetry of
the EH action that one can construct by modification of basic DIs is also a
symmetry of the ADM action, and \textit{vice versa}. This can be demonstrated
in general because any DI written for one field parametrization can be
rewritten for another. Note that a plurality of symmetries in one
parametrization can be eliminated if we restrict our choice to the symmetries
that possess group properties; but for this one particular symmetry there is
still a plurality of parametrization choices.

The modification of any DI under the change of field variables is a general
procedure. According to Noether's theorem, if a transformation is known, then
the corresponding DI can be easily found. If the variables are changed, then
the same changes should appear in the corresponding DIs. We shall briefly
describe this procedure using the ADM variables as an illustrative example,
and one particular DI that is responsible for diffeomorphism.

Let us consider%

\begin{equation}
L_{EH}\left(  g_{\mu\nu}\right)  \rightarrow L_{Dirac}\left(  g_{\mu\nu
}\right)  \rightarrow L_{ADM}\left(  g_{\mu\nu}\left(  N,N^{i},g_{km}\right)
\right)  . \label{eqn110}%
\end{equation}
If we know the DI for the original formulation (the EH action)%

\begin{equation}
I_{\alpha}=-2\left(  g^{\mu\nu}E_{\mu\alpha}\right)  _{,\nu}-g_{,\alpha}%
^{\mu\nu}E_{\mu\nu}\equiv0, \label{eqn111}%
\end{equation}
then we can determine how this DI transforms under a change of variables. (We
use the identity with a covariant index $I_{\alpha}$, instead of $I^{\alpha}$
from (\ref{eqn48}), because of the specific form of the ADM change of field
variables, see (\ref{eqn10}).)

We express (\ref{eqn111}) in terms of the new fields and ELDs that follow from
the new Lagrangian. Variations and ELDs are simply connected (the chain rule).
If we want the contravariant parameter $\xi^{\mu}$, we must seek the relation
of the new ELDs $E,E_{i}~$, and $E_{ADM}^{km}$ with $E_{\mu\nu}$ of the
original action and corresponding DIs, i.e. perform a variation with respect
to a contravariant metric,%

\begin{equation}
E=\frac{\delta L}{\delta N}=\frac{\delta L}{\delta g^{\mu\nu}}\frac{\delta
g^{\mu\nu}}{\delta N}=E_{\mu\nu}\frac{\delta g^{\mu\nu}}{\delta N}%
=2E_{00}\frac{1}{N^{3}}-4E_{0k}\frac{N^{k}}{N^{3}}+2E_{pq}\frac{N^{p}N^{q}%
}{N^{3}}, \label{eqn112}%
\end{equation}

\begin{equation}
E_{i}=\frac{\delta L}{\delta N^{i}}=\frac{\delta L}{\delta g^{\mu\nu}}%
\frac{\delta g^{\mu\nu}}{\delta N^{i}}=E_{\mu\nu}\frac{\delta g^{\mu\nu}%
}{\delta N^{i}}=+2E_{0k}\frac{1}{N^{2}}-2E_{kq}\frac{N^{q}}{N^{2}},
\label{eqn113}%
\end{equation}

\begin{equation}
E_{ADM}^{km}=\frac{\delta L}{\delta g_{km}}=\frac{\delta L}{\delta g^{\mu\nu}%
}\frac{\delta g^{\mu\nu}}{\delta g_{km}}=E_{\mu\nu}\frac{\delta g^{\mu\nu}%
}{\delta g_{km}}=-E_{pq}e^{pk}e^{qm}. \label{eqn114}%
\end{equation}
Solving these equations for the ELDs of the original formulation (ELDs of the
EH action) gives%

\begin{equation}
E_{pq}=-g_{pk}g_{qm}E_{ADM}^{km}~, \label{eqn115}%
\end{equation}

\begin{equation}
E_{0k}=\frac{1}{2}N^{2}E_{k}-N^{q}g_{pk}g_{qm}E_{ADM}^{pm}~, \label{eqn116}%
\end{equation}
and%

\begin{equation}
E_{00}=\frac{1}{2}N^{3}E+N^{2}N^{k}E_{k}-N^{q}N^{k}g_{pk}g_{qm}E_{ADM}^{pm}~.
\label{eqn117}%
\end{equation}

Substituting equations (\ref{eqn115})-(\ref{eqn117}) into (\ref{eqn111}), and
also by expressing the contravariant metric in terms of the ADM variables
using (\ref{eqn10}), the same DIs (\ref{eqn80})-(\ref{eqn81}) follow.
Therefore, if a DI is known we can easily find a new one based on a known
change of variables; in particular, one may find the DI that describes the
canonical (with group property) symmetry of the EH action: diffeomorphism.

Note that if the Lagrangian method is to be considered an
algorithm\footnote{Exactly as Hamiltonian methods are often presented, e.g.
\cite{Horava}: \textquotedblleft one of the advantages of the Hamiltonian
formulation is that one does not have to specify the gauge symmetries a
priori. Instead, the structure of the Hamiltonian constraints provides an
essentially algorithmic way in which the correct gauge symmetry structure is
determined automatically\textquotedblright.} for finding an \textit{a priori}
unknown gauge invariance, then one has to develop some procedure to build DIs
from a given set of ELDs. For covariant theories, where a covariant result
should be expected for the DIs, there is not much flexibility in constructing
the DIs; among the plurality of possible DIs, we should choose (almost
uniquely) DIs such as (\ref{eqn48}) or (\ref{eqn111}) for the EH action. But
it is not clear to us how to formulate an algorithm that, for example for the
ADM Lagrangian, uniquely chooses DIs such as (\ref{eqn80})-(\ref{eqn81})
instead of the DIs for the ADM transformations (\ref{eqn46})-(\ref{eqn47})
which are simpler. What procedure and what criteria could one possibly use to
select such DIs, among the plurality of DIs, without the structure and
guidance from covariance?

Is it acceptable to have this plurality of parametrizations (ADM is only one
of the many possible) and are they all equally good? Alternatively, as in the
case of changing the DI in a particular parametrization, can some condition be
found that allows us to choose a unique parametrization? In the original,
natural parametrization of the EH action, which supports manifest covariance,
the DI that corresponds to diffeomorphism is a covariant expression; both DIs,
(\ref{eqn48}) and (\ref{eqn111}), are true vectors, and if a true vector
equals zero in one coordinate system then it is zero in all coordinate
systems. All non-covariant parametrizations convert these DIs into
non-covariant expressions, which are not true vectors or tensors, and cannot
be zero in all coordinate systems. Only by imposing additional restrictions on
possible coordinate transformations can these non-covariant DIs keep their
form. This limitation explains the origin of yet another name for gauge
invariance in the ADM formulation, \textquotedblleft foliation preserving
diffeomorphism\textquotedblright,\ which reflects the nature of these severe
restrictions. Only changes of coordinates that preserve foliation are allowed:
i.e. space-like surfaces go to space-like surfaces (see \cite{KK-Annals} and
references therein). It should be possible to explicitly relate the form of
DIs in different parametrizations to the restrictions on coordinate
transformations that they impose. So, only one original parametrization leads
to a gauge invariance that is independent of a change of coordinates. The
preservation of covariance in covariant gauge theory must be adopted as one of
the criteria for avoiding plurality.

Consider the Dirac Hamiltonian method; why does it pick only one symmetry in
one parametrization, and a different symmetry in another? How does this
method, not being covariant in nature (time is singled out), successfully
select the unique symmetry and parametrization for a given action? Suppose we
do not know \textit{a priori} about the covariance/parametrization of an
action or its gauge symmetries. When Dirac's method is applied to the EH
action in its original, metric form, despite that time plays a special role in
Hamiltonian methods, covariance is not destroyed because at the end of the
Dirac procedure the transformations of the Lagrangian can be restored; they
are covariant and exactly the same as those known from the Lagrangian approach
\cite{KKRV, Myths}. Applying the same method (Dirac's) to the ADM Lagrangian
(another parametrization) also gives a symmetry -- the ADM symmetry, which
does not preserve the covariant form and does not have group properties.
Dirac's method can be used to select the parametrization (that can be called
\textquotedblleft canonical\textquotedblright) in which the Lagrangian is
written in a natural form, and in which the \textquotedblleft
canonical\textquotedblright\ Hamiltonian leads to the symmetry with group
properties. Therefore, the advantage of the Dirac method lies in its
field-parametrization dependence, which compensates for the lack of manifest
covariance; and it is only because of its parametrization dependence that it
can be useful in finding canonical variables, canonical symmetry, \textit{et
cetera}.

Hamiltonian methods are very sensitive to the choice of parametrization in
covariant theories; this is clear from the effect that a non-covariant
parametrization has on DIs, which after non-covariant changes retain their
form only under additional restrictions on the changes of coordinate system.
That is why it is important to choose the right parametrization before the
Hamiltonian method is applied. The answer to the question posed by Shestakova
in \cite{ShestakovaCQG}: \textquotedblleft Would not it be better to restrict
ourself by transformations in phase space of original canonical variables in
the sense of Dirac?\textquotedblright\ is \textquotedblleft
yes\textquotedblright. Covariant theories (all fundamental physical
interactions are described by such theories) are built on the fundamental
physical principle -- covariance; the gauge symmetries related to them are
also expressed in covariant form, and of course the use of natural variables
in which the covariance is manifest is preferable. But Shestakova made a
different choice; she concluded that Dirac's method is \textquotedblleft not
fundamental and undoubted\textquotedblright\ because it is
parametrization-dependent, and it prevents one from obtaining diffeomorphism
invariance for the ADM parametrization, which is considered important because
of its \textquotedblleft geometrical interpretation\textquotedblright%
\ \cite{ShestakovaCQG}. But this interpretation is related to
\textquotedblleft foliation preserving diffeomorphism\textquotedblright\ and
the restriction on coordinate transformations; this is exactly the
interpretation that Hawking stated \textquotedblleft to be contrary to the
whole spirit of relativity\textquotedblright\ \cite{Hawking}. The true puzzle
is why one would need to dismiss the Dirac method (and covariance of the EH
action) as not fundamental, and to search for \textquotedblleft a clear
proof\textquotedblright\ that can \textquotedblleft restore a legitimate
status of the ADM parametrization\textquotedblright\ \cite{ShestakovaCQG}.

The parametrization dependence of Dirac's method is not limited to
compensating for the lack of manifest covariance in his approach. It equally
well plays an important role in non-covariant models, where it also allows one
to find the canonical or natural parametrization of the gauge-invariant
Lagrangian; in these variables, the corresponding symmetry has the simplest
commutator \cite{KKK-2} (examples that explicitly illustrate this point and
descriptions how to find such parametrizations will be reported elsewhere
\cite{KKK-4}).

One additional advantage of Dirac's method can be illustrated by the example
of the EH action, where even imposing a requirement to use covariant DIs at
the Lagrangian level leaves some freedom. One can build some additional
covariant DIs, which are obviously satisfied, for example (there are still
only four independent DIs),%

\begin{equation}
D_{\nu}D^{\nu}I^{\mu}\equiv0\text{ \ \ \ \ \ or \ \ \ }D_{\nu}D^{\mu}I^{\nu
}\equiv0, \label{eqn120}%
\end{equation}
and their corresponding transformations can be found. In the Hamiltonian
analysis, however, it is not possible to obtain such transformations since it
always picks the simplest one with the lowest possible order of derivatives of
the ELDs in the DIs. The order of derivatives is related to the degrees of
freedom (DOF) counting in the Hamiltonian formulation for constrained systems
(e.g. see \cite{Henneaux}), which is based on the constraint structure of the
Hamiltonian formulations: the length of constraint chains is related to the
order of derivatives of the gauge parameters presented in the gauge
transformation and the number of primary first-class constraints equals the
number of gauge parameters. For DIs such as (\ref{eqn120}), there should be
constraints up to fourth-order, and four gauge parameters, i.e. minus 16 DOF
for only ten components of metric; this is a negative number and, therefore,
not a physical result.

\section{Conclusion}

Our original intention for writing the present paper was to provide the
solution to the \textquotedblleft non-canonicity puzzle\textquotedblright%
\ described in \cite{CLM}. Using the Lagrangian method, we have already shown
in \cite{KKK-1} that the EH action is invariant under both transformations --
diffeomorphism and the one that follows from the Hamiltonian formulation of
the ADM gravity. The same was also confirmed for the ADM action: it is
invariant under both these transformations. Therefore diffeomorphism is not
lost in the ADM action at the Lagrangian level; yet, this is not the solution
to the \textquotedblleft puzzle\textquotedblright\ because these two
formulations (EH and ADM) are not equivalent at the Hamiltonian level, and the
Dirac analysis leads to unique, though different gauge symmetries for these
two formulations.

What is a possible solution to the \textquotedblleft puzzle\textquotedblright?
One solution would be to modify the Dirac procedure and force it to produce
the \textquotedblleft expected symmetry\textquotedblright\ for all conceivable
parametrizations. Another solution would be to respect the plurality of gauge
symmetries at the Lagrangian level, and to try to relate each symmetry at the
Lagrangian level to a particular parametrization that leads uniquely to this
symmetry at the Hamiltonian level (i.e. do not modify the Dirac procedure).
Returning to the epigraph of our paper: \textquotedblleft Plurality must never
be posited without necessity\textquotedblright, we are faced with another
question: are all possible parametrizations and DIs equally good? Among the
plurality of parametrizations and gauge symmetries (linear combinations of
DIs), there is one Lagrangian symmetry that possesses a group property --
diffeomorphism; and it is exactly this symmetry that follows naturally from
the original Einstein formulation of General Relativity. 

If we start from the EH action%

\begin{equation}
S_{EH}=\int\sqrt{-g}Rd^{4}x~, \label{eqn121}%
\end{equation}
for which the ELD is%

\begin{equation}
E^{\alpha\beta}=\frac{\delta L_{EH}}{\delta g_{\alpha\beta}}=\sqrt{-g}\left(
\frac{1}{2}g^{\alpha\beta}R-R^{\alpha\beta}\right)  , \label{eqn122}%
\end{equation}
then it leads to the DI%

\begin{equation}
I^{\mu}=\nabla_{\nu}E^{\mu\nu}\equiv0 \label{eqn123}%
\end{equation}
and the corresponding gauge transformation:%

\begin{equation}
\delta g_{\mu\nu}=-\nabla_{\mu}\xi_{\nu}-\nabla_{\nu}\xi_{\mu}~.\label{eqn124}%
\end{equation}
Note, all these expressions are generally covariant. In natural
parametrization, metric tensor, the Dirac procedure leads exactly to the same
symmetry at the Hamiltonian level \cite{Myths, KKRV}. All other field
parametrizations (e.g. ADM) can be used to rewrite (\ref{eqn122}) -
(\ref{eqn124}), but general covariance would be lost. In addition, the Dirac
procedure would produce different symmetries for different filed parametrizations.

When delivering the 1933 Herbert Spencer lecture, Einstein said:
\textquotedblleft It is the grand object of all theory to make these
irreducible elements as simple and as few in number as possible, without
having to renounce the adequate representation of any empirical content
whatever\textquotedblright, and later: \textquotedblleft Our experience
hitherto justifies us in believing that nature is the realization of the
simplest conceivable mathematical ideas\textquotedblright%
\ \cite{Einstein-ideas}. What could be simpler than equations (\ref{eqn121}%
)-(\ref{eqn124})? Do we need another set of field variables that will
complicate them and destroy covariance?

There is no \textquotedblleft non-canonicity puzzle\textquotedblright,\ and
there is no \textquotedblleft contradiction that again witnesses about the
incompleteness of the theoretical foundation\textquotedblright%
\ \cite{ShestakovaCQG}, unless one desperately wants to find \textquotedblleft
a clear proof\textquotedblright\ of the legitimacy of the non-covariant ADM
variables because these variables are a \textquotedblleft common
currency\textquotedblright\ \cite{Pullin}. In our opinion, to support the
legitimacy of the common currency and its derivatives may incur a high cost;
indeed, one might barter away valuable physical assets.

\section{Acknowledgment}

We would like to thank A.M. Frolov, L.A. Komorowski, D.G.C. McKeon, and A.V.
Zvelindovsky for discussions.


\begin{thebibliography}{99}                                                                                               %


\bibitem {CLM}F. Cianfrani, M. Lulli, G. Montani, arXiv:1104.0140 [gr-qc]

\bibitem {Diracbook}P.A.M. Dirac, \textit{Lectures on Quantum Mechanics}
(Belfer Graduate School of Sciences, Yeshiva University, New York, 1964)

\bibitem {ADM}R. Arnowitt, S. Deser, C.W. Misner, in \textit{Gravitation: An
Introduction to Current Research}, ed. by L. Witten (Wiley, New York, 1962),
p. 227; arXiv:gr-qc/0405109

\bibitem {goldies}R. Arnowitt, S, Deser, C.W. Misner, Gen. Relativ. Gravit.
\textbf{40}, 1997 (2008)

\bibitem {Myths}N. Kiriushcheva, S.V. Kuzmin, Central Eur. J. Phys.
\textbf{9}, 576 (2011)

\bibitem {PSS}F.A.E. Pirani, A. Schild, R. Skinner, Phys. Rev. \textbf{87,}
452 (1952)

\bibitem {Dirac}P.A.M. Dirac, Proc. Roy. Soc. A \textbf{246,} 333 (1958)

\bibitem {Rovelli}C. Rovelli,\textit{ Quantum Gravity} (Cambridge University
Press, Cambridge, 2004)

\bibitem {KKRV}N. Kiriushcheva, S.V. Kuzmin, C. Racknor, S.R. Valluri, Phys.
Lett. A \textbf{372}, 5101 (2008)

\bibitem {FKK}A.M. Frolov, N. Kiriushcheva, S.V. Kuzmin, S.V.: arXiv:0809.1198
[gr-qc] (to appear in Grav. \& Cosm., \textbf{17}, No 4, 2011)

\bibitem {Lanczosbook}C. Lanczos, \textit{The variational principles of
mechanics}, fourth ed. (Dover Publications, New York, 1970)

\bibitem {Isham}C.J. Isham, K.V. Kuchar, Ann. Phys. \textbf{164}, 316 (1985)

\bibitem {KKK-1}N. Kiriushcheva, P.G. Komorowski, S.V. Kuzmin, arXiv:1107.2449 [gr-qc]

\bibitem {Noether}E. Noether, Nachr. d. K\"{o}nig. Gesellsch. d. Wiss. zu
G\"{o}ttingen, Math.-phys. Klasse, \textbf{2}, 235 (1918)

\bibitem {Noether-eng}E. Noether (M.A. Tavel's English translation), arXiv:physics/0503066

\bibitem {Landau}L.D. Landau, E.M. Lifshitz, \textit{The Classical Theory of
Fields}, fourth ed. (Pergamon Press, Oxford, 1975)

\bibitem {Carmeli}M. Carmeli, \textit{Classical Fields, General Relativity and
Gauge Theory} (World Scientific, New Jersey, 2001)

\bibitem {Ostr}M. Ostrogradsky, Memoires l'Acad. Imperiale Sci.
St.-Petersbourg, \textbf{IV}, 385 (1850)

\bibitem {Ostr-rus}M. Ostrogradsky, in: \textit{Variatsionnye printsipy
mekhaniki}, ed. by L.S. Polak (Fizmatgiz, 1959), p. 315

\bibitem {Whittaker}E.T. Whittaker, \textit{A Treatise on the Analytical
Dynamics of Particles and Rigid Bodies} (Dover Publications, 1944)

\bibitem {GLT}D. M. Gitman, S.L. Lyakhovich and I. V. Tyutin, Izvestiya Vuz.
Fiz. \textbf{26}, 61 (1983); Sov. Phys. J. \textbf{26}, 730 (1984)

\bibitem {GTbook}D. M. Gitman and I. V. Tyutin, \textit{Quantization of Fields
with Constraints} (Springer, Berlin, 1990)

\bibitem {DD}S.K. Dutt and M. Dresden, \textit{Pure gravity as a constrainted
second-order system}, Preprint ITP-SB-86-32\ (1986)

\bibitem {2D}R.N. Ghalati, N. Kiriushcheva and S.V. Kuzmin, Mod. Phys. Lett.
\textbf{A 22}, 17 (2007)

\bibitem {Castellani}L. Castellani, Ann. Phys. \textbf{143}, 357 (1982)

\bibitem {Pons2003}J.M. Pons, Class. Quantum Grav. \textbf{20}, 3279 (2003)

\bibitem {Gravitation}C.W. Misner, K.S. Thorne, J.A. Wheeler,
\textit{Gravitation} (W.H. Freeman and Company, San Francisco, 1973)

\bibitem {Waldbook}R.M. Wald, \textit{General Relativity} (The University of
Chicago Press, Chicago, 1984)

\bibitem {Horava}P. Ho\v{r}ava and C.M. Melby-Thompson, Phys. Rev. \textbf{D
82}, 064027 (2010)

\bibitem {Pullin}J. Pullin, \textit{Golden Oldie Editorial}, Gen. Relativ.
Gravit. \textbf{40}, 1989 (2008)

\bibitem {Einstein1925}A. Einstein, Sitzungsber. preuss. Akad. Wiss.,
Phys.-Math. \textbf{1}, 414 (1925)

\bibitem {Einstein1925-rus}A. Einstein, \textit{The Complete Collection of
Scientific Papers}, vol. \textbf{2} (Nauka, Moscow, 1966), p. 171

\bibitem {Einstein-eng}English translation of some of Einstein's papers is
available from: http:/www.lrz-muenchen.de/`aunzicker/ae1930.htm, and A.
Unzicker, T. Case, arXiv:physics/0503046

\bibitem {KK-Annals}N. Kiriushcheva and S. V. Kuzmin, Ann. Phys. \textbf{321}
(2006) 958

\bibitem {affine-metric}N. Kiriushcheva, S.V. Kuzmin, Eur. Phys. J. \textbf{C}
\textbf{70}, 389 (2010)

\bibitem {G/R}R.N. Ghalati and D.G.C. McKeon, arXiv:0712.2861v3 [gr-qc]

\bibitem {Gerry}D.G.C. McKeon, Int. J. Mod. Phys. \textbf{A 25}, 3453 (2010)

\bibitem {Schwinger}J. Schwinger, Phys. Rev. \textbf{130}, 1253 (1963)

\bibitem {Hilbert}D. Hilbert, Nachr. d. K\"{o}nig. Gesellsch. d. Wiss. zu
G\"{o}ttingen, Math.-phys. Klasse, \textbf{8}, 395 (1916)

\bibitem {Hilbert-eng}D. Hilbert, in \textit{The Genesis of General
Relativity}, ed. by J. Renn (Springer, 2007) vol. \textbf{4}, p. 1003

\bibitem {Trans}N. Kiriushcheva, S.V. Kuzmin, Gen. Rel. Grav. \textbf{42},
2613 (2010)

\bibitem {Pullin-18}J. Pullin, AIP Conf. Proc. \textbf{668}, 141 (2003), arXiv:gr-qc/0209008

\bibitem {Higgs}P.W. Higgs, Phys. Rev. Lett. \textbf{1}, 373 (1958)

\bibitem {Erratum}P.W. Higgs, Phys. Rev. Lett. \textbf{3}, 66 (1959)

\bibitem {DeWitt}B.S. DeWitt, Phys. Rev. \textbf{160}, 1113 (1967)

\bibitem {Shestakova2004}T.P. Shestakova and C. Simeone, Grav. \& Cosm.
\textbf{10}, 161 (2004)

\bibitem {ShestakovaCQG}T.P. Shestakova, Class. Quantum Grav. \textbf{28},
055009 (2011)

\bibitem {Einstein-Science}A. Einstein, Science, \textbf{74}, 438 (1922)

\bibitem {Kuchar}K. Kucha\v{r}, in: \textit{Relativity, Astrophysics and
Cosmology}, ed. by W. Israel (D. Reidel Publ. Comp, Dordrecht, 1973), p. 237

\bibitem {Haw-Pen}S.W. Hawking, R. Penrose, Proc. Roy. Soc. \textbf{A 314},
529 (1970)

\bibitem {Hawking}S.W. Hawking, in: \textit{General Relativity. An Eistein
Centenary Survey}, ed. by S.W. Hawking and W. Israel (Cambridge University
Press, Cambridge, 1979), p. 746

\bibitem {KKK-2}N. Kiriushcheva, P.G. Komorowski, S.V. Kuzmin, arXiv:1107.2981 [gr-qc]

\bibitem {KKK-4}N. Kiriushcheva, P.G. Komorowski, S.V. Kuzmin (in preparation)

\bibitem {Henneaux}M. Henneaux, C. Teitelboim and J. Zanelli, Nucl. Phys.
\textbf{B} \textbf{332}, 169 (1990)

\bibitem {Einstein-ideas}A. Einstein, \textit{Ideas and Opinions}, ed. by C.
Weltbild, translated by S. Bargmann (Wings Books, New York, 1954)
\end{thebibliography}
\end{document}